\documentclass[prl,twocolumn,showpacs,preprintnumbers,amsmath,amssymb]{revtex4}
\usepackage{epsfig}
\usepackage{wasysym}
\usepackage{amssymb}
\usepackage{slashed}
\usepackage{physics}
\usepackage{color}

\begin{document}

\newcommand{\bvec}[1]{\mbox{\boldmath ${#1}$}}
\title{Three Dimensional Black Hole in the Low Energy Heterotic String Theory}
\author{Byon N. Jayawiguna}\email[]{byon.nugraha@.ui.ac.id}
\affiliation{Departemen Fisika, FMIPA, Universitas Indonesia, Depok 16424, Indonesia}
\date{\today}
\begin{abstract}
 We study the BTZ black holes (2+1 dimensional space-time) in the low energy heterotic string theory (BTZ-Sen BH). This concept requires us to include a non-trivial dilaton $ \phi $ and a 3-form $ H _ {\mu \nu \rho} $ field. By using the Hassan-Sen transformation and BTZ black hole as a seed solution, we obtain the solution in the string frame. Some properties of the black hole solutions are discussed.
\end{abstract}
\pacs{97.60.Lf, 04.20.Jb}

\maketitle

\section{Introduction}
Banados, Teitelboim, and Zanelli (BTZ) in 1992 were the first to obtain exact solution of a vacuum rotating black hole in three-dimensional space-time with non-zero cosmological constant \cite{Banados:1992wn, Banados:1992gq}. This proposal also supports the existence of three-dimensional space-time in general relativity as the foundation of the classical and quantum gravity aspects that have been conceived by Deser \cite{Deser:1983tn}, Jackiw \cite{Deser:1988qn}, 't Hooft \cite{tHooft:1988qqn}, and Witten \cite{Witten:1988hc, Witten:1989sx} circa 1984-1989. Because the three-dimensional space-time has no Newtonian limit, BTZ black holes are different from Kerr black holes \cite{Kerr:1963ud}. The fundamental difference is that the BTZ black holes are asymptotically anti-de-Sitter and do not have a curvature singularity, whereas the Kerr black holes are asymptotically flat and have a ring singularity. Nevertheless, both objects are still black holes; have the inner horizon as well as the outer horizon, ergopshere, static limit region, and frame dragging effects.

From the vacuum BTZ case, it took eight years for Martinez, Teitelboim, and Zanelli to obtain the charged and rotating black holes in three-dimensional space-time \cite{Martinez:1999qi}. The method used are the same with BTZ. They do not neglect the boundary action, with the consequence that the scalar Ricci evaluated on the hypersurface $ \Sigma $ must be taken into account. With the Hamiltonian formulation and $(2+1)$ decomposition, they choose stationary conditions and axial symmetry for the consequences of rotating object. Under these conditions, the lapse and shift functions, $N^{\perp}$ and $N^{i}$, could be obtained using the Hamiltonian and momentum constraints $ \mathcal{H}_{\perp}=0 $ and $ \mathcal{H}_{i}= 0 $.

Although the BTZ black hole solution is a rotating object, it is without doubt that there is a vast literature on the static BTZ black hole $ (\Lambda \neq 0) $. BTZ with radial and azimuthal electric fields has been studied by the authors in~\cite{Martinez:1999qi,Cataldo:2002fh, Hirschmann:1995he} while BTZ with pure magnetic solutions has also been studied in~\cite{Koikawa:1997am, Cataldo:2000ns}. Interestingly, studies on static BTZ black holes have been widely discussed in the case of nonlinear electrodynamics as well as the phenomenological aspects \cite{Myung:2008kd, Hendi:2017oka, Dehghani:2019noj, Fernando:2017qrq, Balart:2019uok, Fernando:2019cez, Gurtug:2018bxu}.

In the realm of low energy string theory, Sen \cite{Sen:1992ua} obtained an exact four-dimensional, rotating and charged black hole solution, commonly referred to as a Kerr-Sen black hole. This solution is obtained by the Hassan-Sen transformation \cite{Hassan:1991mq} which requires a seed solution (in the literature, Sen uses a Kerr black hole solution) that satisfies the stationary and axial symmetry conditions. The black hole solution obtained by Sen is analogous to the Kerr-Newman black hole (rotating charged black hole in four dimensional space-time). Despite the similarities between the Kerr-Sen and Kerr-Newman black hole solutions, there are several features that could distinguish between the two solutions. For example, the appearence of dilaton scalar and 3-form tensor field on Kerr-Sen. It can be inferred that the Kerr-Sen and Kerr-Newman black hole is the same but different families. On the other hand, a study on the merger estimates \cite{Siahaan:2019oik} and the energy requirements needed to destroy the horizon of the Kerr-Sen black hole \cite{Siahaan:2015ljs} shows that the study of Kerr-Sen black holes is still of particular interest for researchers. Moreover, since the black hole in string theory is consistent with the descriptions of classical and quantum gravity, a vast literature on AdS/CFT correspondence to the Kerr-Sen black hole are discussed in the reference \cite{Ghezelbash:2012qn, Siahaan:2015xia, Siahaan:2018wvh}. Recently, accelerating black hole in the low energy string theory and several aspects such as horizons, conical singularities, angular velocity, and temperatures have been obtained \cite{Siahaan:2018qcw}.

For the case of static black holes in string theory, Gibbons and Maeda have obtained black holes in extra dimensions \cite{Gibbons:1987ps} (also Horowitz and Strominger \cite {Horowitz:1991cd}) while in the four dimension Garfinkle, Horowitz, and Strominger have studied the charged black hole in string theory (Einstein-Maxwell-Dilaton) \cite{Garfinkle:1990qj}. On the other hand, the Kerr-Sen black hole is parameterized with a mass $ M $, charge $ Q $, and angular momentum $ J $. If we set $ Q = 0 $, the solution will be reduced to Kerr black hole solution. If we switch off the contribution of the rotation parameter $ a $ followed by the transformation of $ r \rightarrow r - \frac{Q^2}{M} $ \cite{Siahaan:2014ihe}, we can find that the solution is nothing but a static charged black hole solution in string theory \cite{Garfinkle:1990qj}. It can be inferred that in order to arrange the static and charged string object, a coordinate transformation must be performed.

In this paper, we devote our effort to constructing what Sen had already obtained on Kerr-Sen black hole to the BTZ black hole object. The output of this paper is to obtain the BTZ black hole in the low energy heterotic string theory. We also include the contribution of cosmological constant $ \Lambda = -\frac{1}{l^2} $ since the BTZ black hole will exist if it has a negative cosmological constant (anti-de-Sitter spacetime). BTZ black holes in both rotating and static in low energy string field have been studied extensively in some literature \cite{Horowitz:1993jc, Horowitz:1994ei, Chan:1994qa, Chan:1995wj, Chan:1996rd, Sa:1995vs,Garcia-Diaz:2017cpv,Aniceto:2017gtx,Fernando:2009tv,Fernando:2004ay,Fernando:2003ai}. However, construction of a rotating, charged black hole with a complete low energy string term $(R,~\Lambda,~F_{\mu\nu},~\phi,~H_{\mu \nu \rho}) $ has never been studied.

The paper is structured as follows. In section II we will do a quick review on BTZ vacuum black hole. This is used because the incoming seed solution for Hassan-Sen transformation is a BTZ black hole solution. This section will be filled by the metric solution, event horizons, and static limit radius. In section III, we construct the BTZ black hole on the low energy heterotic strings (BTZ-Sen). From the solution we obtained in the string frame, we transform conformally into Einstein frame. After we obtain the solution, we also discuss in detail this solution to study the geometrical aspects of the BTZ-Sen black hole (event horizon, static limit radius, and ergosphere) and the test particles inside and outside of the black hole. The angular velocity of the particle and black hole perimeter are also discussed.

\section{BTZ Black Hole: A Review}

Three dimensional (2 + 1) black holes or commonly known as BTZ black holes were first obtained in 1992 \cite{Banados:1992gq, Banados:1992wn} (see also \cite{Carlip:1995qv}). This black hole is quite easy to imagine compared to the black hole that we are familiar with (Schwarzschild, Kerr, etc.) since the BTZ black hole has a lower dimension than the ordinary four dimension (3+1).  For more details on the concept above, the author begins with the action form of the BTZ black hole. The BTZ black hole action is given by
\begin{equation}
I=\frac{1}{16 \pi G} \int \sqrt{-g} \left(R+2 l^{-2}\right) d^3 x,
\end{equation}
from the action above, the symbol $ g $ is stands for determinant of the metric tensor $ g_{\mu\nu} $ whereas $ R $ and $ l $ is the Ricci scalar and the radius associated with the cosmological constant $ -\Lambda = l^{-2} $, respectively. The corresponding Einstein's equation is
\begin{equation}
R_{\mu\nu}-\frac{1}{2}g_{\mu\nu} R + \Lambda g_{\mu\nu}=0.
\end{equation}
The equation above can be made in the nice way by contracting it with $ g^{\mu\nu} $. Then we found the equation to be
\begin{equation}
\label{persamaan}
R_{\mu\nu}=2\Lambda g_{\mu\nu}.
\end{equation} 
In general, if $ T_{\mu\nu} \neq 0 $, the Einstein field equation can be written as
\begin{equation}
\label{persamaanT}
R_{\mu\nu}=2\Lambda g_{\mu\nu} + 8\pi G (T_{\mu\nu}-g_{\mu\nu}T),
\end{equation}
In addition, the selection of the momentum energy tensor $ T_{\mu\nu} $ completely determines the Ricci $ R_{\mu\nu} $ tensor, but generally does not determine the Riemann tensor $ R_{\mu\nu\alpha\sigma} $. However, in the 3-dimensional gravity concept, the traceless part of $ R_{\mu\nu \alpha \sigma} $ vanishes so that the Riemann tensor only depends linearly with its Ricci scalar and Ricci tensors
\begin{eqnarray}
R_{\mu\nu\gamma\sigma}&=& g_{\mu\gamma}R_{\nu\sigma}+g_{\nu\sigma}R_{\mu\gamma}-g_{\nu\gamma}R_{\mu\sigma}-g_{\mu\sigma}R_{\nu\gamma}\nonumber \\ && -\frac{1}{2} (g_{\mu\gamma}g_{\nu\sigma}-g_{\mu\sigma}g_{\nu\gamma}) R.
\end{eqnarray} 
By inserting Eq.~\ref{persamaanT}, we get a relationship between the Riemann tensor and the momentum energy tensor. For vacuum cases $ (T_{\mu\nu}=0) $, the Riemann tensor and Ricci scalar can be written as $ R_{\mu\nu\gamma\sigma}= (g_{\mu\gamma}g_{\nu\sigma}-g_{\mu\sigma}g_{\nu\gamma})\Lambda $, and $ R=6\Lambda $.
Eq.~\ref{persamaan} above can be solved to obtain the black hole solution with ansatz
\begin{equation}
ds^2 = g_{\mu\nu}dx^{\mu}dx^{\nu}= -N^2 dt^2 + N^{-2} dr^2 +r^2 \left(N^{\phi} dt + d\phi\right)^2,
\end{equation}
Where $ N(r) $ and $ N^{\phi}(r) $ are the lapse and the angular shift function, respectively. The solution is written as
\begin{eqnarray}
\label{solusi}
N(r)= -M +\frac{r^2}{l^2}+\frac{J^2}{4r^2},~ N^{\phi}(r)= -\frac{J}{2r^2},
\end{eqnarray}
with $ -\infty<t<\infty$, $~0<r<\infty $, and $ 0\leq \phi \leq 2\pi $. The metric solution could be written in a form
\begin{eqnarray}
ds^2 &=& -\left(\frac{r^2}{l^2}-M\right) dt^2  - J dtd\phi+ \left(\frac{r^2}{l^2}-M+\frac{J^2}{4r^2} \right)^{-1} dr^2 \nonumber \\ && +r^2 d\phi^2.
\end{eqnarray} 
The two constants $ M $ and $ J $ contained in Eq.~\eqref{solusi} are the consequences of stationary conditions (invariant with respect to time) and axial symmetry (invariant with respect to rotation), mass and angular momentum. In addition, the metric $ g^{rr} $ will be zero if its obey the value of $ r $ as follows
\begin{equation}
r_{\pm} = l \sqrt{\frac{M}{2} \left[ 1\pm \sqrt{1-\left(\frac{J}{M l}\right)^2}   \right]}.
\end{equation}  
The value of $ r $ indicates that there are two event horizon in the case of the BTZ black hole. The three variables $ l $, $ M $, and $ J $ play a major role in this case. To keep the event horizon on this BTZ black hole exist, the variables $ l $, $ M $, and $ J $ must satisfy the relation below
\begin{eqnarray}
M>0,~ |J|\leq Ml.
\end{eqnarray}
In extreme conditions $ |J| = Ml $, event horizon $ r_{+} $ $ r_{-} $ merges into one with a radius $ r_{e} = l \sqrt{\frac{M}{2}} $. Another aspect that can be obtained is the emergence of ergosphere
\begin{equation}
r_{sl}= l\sqrt{M}.
\end{equation}
With all these $ r $ values obtained, it can be said that the characteristics of BTZ black holes are almost the same as those of Kerr black holes, $ r_{+} $ is an event horizon, $ r_{sl} $ is a static limit, and the range of $ r_{+} <r <r_{sl} $ is the ergosphere area.

\section{BTZ Black Hole in the low energy Heterotic String Theory}
Black hole in string field (such as \textit{black p-brane}) certainly has different characteristics compared to black holes that often appear in Einstein's theory of gravity. Most of these solutions contain more than one charge such as the Yang-Mills field or the antisymmetric tensor field, and the dilaton scalar field. When these charges are switched off, the solution will be reduced to an ordinary Schwarzschild solution. In this paper, we construct the classical and exact solution of the BTZ black hole in the low energy string that describe black hole carrying some charge and angular momentum. The method used to obtain this solution is the twisting procedure which requires a stationary and axial symmetry solution as an input. This twisting procedure which will later be referred to as Hassan-Sen transformation. The explanation above starts from the action of string theory in the 3d string frame
\begin{equation}
\label{aksi}
S= \int d^3x \sqrt{-g}~ e^{-\Phi} \left(\mathcal{R}-2\Lambda+ (\nabla\Phi)^2 -\frac{F^2}{8} -\frac{H^2}{12}\right),
\end{equation}
where $ g $ is a determinant of metric tensor $ g_{\mu\nu} $, $ R $ is a Ricci scalar, $ F_{\mu\nu}=\partial_{\mu}A_{\nu}-\partial_{\nu}A_{\mu} $ is a Maxwell field, $ \phi $ as a scalar dilaton, and
\begin{equation}
H_{\kappa\mu\nu}= \partial_{\kappa}B_{\mu\nu} + cyclic~permutation-\left[\Omega_{3}(A)\right]_{\kappa\mu\nu},
\end{equation}
where $ B_{\mu\nu} $ is antisymmetry tensor field, and
\begin{equation}
\left[\Omega_{3}(A)\right]_{\kappa\mu\nu} = \frac{1}{4} \left(A_{\kappa}F_{\mu\nu}~+~ cyclic~permutation\right)
\end{equation}
is a Chern-Simons term. The $3$-form field tensor $ H_{\kappa \mu \nu} $ above can be made in a compact form into
\begin{eqnarray}
H_{\kappa\mu\nu}&=&\partial_{\kappa}B_{\mu\nu}+ \partial_{\nu}B_{\kappa\mu}+\partial_{\mu}B_{\nu\kappa}\nonumber \\ &&-\frac{1}{4}\left(A_{\kappa}F_{\mu\nu}+A_{\nu}F_{\kappa\mu}+A_{\mu}F_{\nu\kappa}\right).
\end{eqnarray}
One can obtain the equation of motion by varying the action \eqref{aksi} with respect to $ \lbrace g_{\mu\nu},~\Phi,~A_{\mu},~B_{\mu\nu} \rbrace$
\begin{eqnarray}
R_{\mu\nu} &=& g_{\mu\nu}\left[2 \Lambda - \nabla^2\Phi +(\nabla\Phi)^2-\frac{1}{8}F^2 - \frac{1}{6}H^2\right]\nonumber \\ && -\nabla_{\mu}\Phi\nabla_{\nu}\Phi+\frac{1}{4}g^{\lambda \rho}F_{\mu\lambda}F_{\nu\rho} + \frac{1}{4} H_{\mu\kappa\sigma}H_{\nu}^{\kappa\sigma},
\end{eqnarray}
\begin{eqnarray}
 \nabla^2\Phi - (\nabla\Phi)^2 -2\Lambda+\frac{F^2}{8}+\frac{H^2}{6}=0, \nonumber \\  \nabla_{\mu}\left(e^{-\Phi}F^{\mu\nu}\right) = \frac{1}{2} F_{\alpha\beta} H^{\nu\alpha\beta},
\end{eqnarray}
and 
\begin{equation}
\nabla_{\mu}\left(e^{-\Phi}H^{\kappa\mu\nu}\right)=0.
\end{equation}
It is shown that the ordinary BTZ equation in \eqref{persamaan} can be obtained by switched off the non-gravitational field. Thus, taking the BTZ black hole as a seed solutions are allowed in string frame. However, rather than considering the solutions only in the string frame, we will make a conformal transformation to the Einstein frame. We therefore define the conformal transformation 
\begin{equation}
\label{conformal}
ds_{E}^2 = e^{-\Phi} \tilde{ds}^2,
\end{equation} 
The solution in the Einstein frame will be used when we discuss some aspects BTZ black holes in heterotic string theory. Hassan and Sen show that the new field $ \lbrace g'_{\mu\nu},~B'_{\mu \nu},~A'_{\mu},~\phi'\rbrace $ will satisfies the equation of motion of \ref{aksi} if the relationship of the new field with the old is 
\begin{eqnarray}
\label{trans}
\mathcal{M}'=\Omega \mathcal{M}\Omega^{T},~\Phi'=\Phi + \ln\sqrt{\frac{g'}{g}}.
\end{eqnarray}
The definition for matrix $ \mathcal{M} $ and $ \Omega $ in the \eqref{trans} are as follows
\begin{equation}
\footnotesize
\label{matriksm}
\mathcal{M} = 
\begin{pmatrix}
(K^{T}-\eta)g^{-1}(K-\eta) & (K^{T}-\eta)g^{-1}(K+\eta) & -(K^{T}-\eta)g^{-1} A \\
(K^{T}+\eta)g^{-1}(K-\eta) & (K^{T}+\eta)g^{-1}(K+\eta) & -(K^{T}+\eta)g^{-1} A \\
-A^{T}g^{-1}(K-\eta) & -A^{T}g^{-1}(K+\eta) & A^{T}g^{-1}A
\end{pmatrix},
\end{equation}
and
\begin{equation}
\Omega = 
\begin{pmatrix}
I_{5x5} & ... & ... \\
... & \cosh\alpha & \sinh\alpha \\
... & \sinh\alpha & \cosh\alpha
\end{pmatrix}.
\end{equation}
The dot sign in the matrix $ \Omega $ indicate that the contents of the matrix are zero whereas $ I_ {5x5} $ is an identity matrix 5x5. The definition inside the matrix $ M $ could be written as
\begin{align}
	K_{\mu\nu} = 
	\begin{pmatrix}
		K_{rr}  & K_{r\phi} & K_{rt} \\
		K_{\phi r} &  K_{\phi\phi} & K_{\phi t} \\
		K_{tr}    & K_{t \phi}  & K_{tt} 
	\end{pmatrix}, ~ \eta_{\mu\nu} = \begin{pmatrix}
		1 & 0 & 0 \\
		0 & 1 & 0  \\
		0 & 0 &-1  \\ 
	\end{pmatrix},
\end{align}
and
\begin{align}
	 g^{\mu\nu}= \begin{pmatrix}
	g^{rr} & g^{r\phi} & g^{rt} \\
	g^{\phi r} & g^{\phi\phi} & g^{\phi t} \\
	g^{tr}   & g^{t \phi}  & g^{tt}
	\end{pmatrix}.
\end{align}
where the matrix $ K_{\mu\nu} $ and $ A_{\mu} $ is given by
\begin{align}
	K_{\mu\nu}=-B_{\mu\nu}-g_{\mu\nu}-\frac{1}{4}A_{\mu}A_{\nu},~ A_{\mu}=\begin{pmatrix}
		A_{r}    \\
		A_{\phi}  \\
		A_{t}
	\end{pmatrix}.
\end{align}

Here, Sen takes a different definition of the matrix; the time component is placed at the lower right end. However, the new and the old fields obtained in the \eqref{trans} equation will satisfy the equation of motion derived from the \eqref{aksi} action. An explanation of the ideas from Sen is as follows. We know that the action of low energy string theory in the three dimensions will be reduced to the Einstein-Hilbert action when all fields $ \lbrace B'_{\mu\nu},~A'_{\mu},~\Phi'\rbrace $ are zero. That is, the vacuum solution of the Einstein field equation is also a solution of the equation of motion derived from action \eqref{aksi}. So, by using the transformation \eqref{trans} we will obtain a new solution with non vanisihing non gravitational fields $ \lbrace B'_{\mu\nu},~A'_{\mu},~\Phi' \rbrace $.
Ansatz metrics that obey stationary and axial symmetry conditions are as follows
\begin{equation}
\label{ansatz}
g_{\mu\nu}dx^{\mu}dx^{\nu}= g_{tt} dt^2+ 2 g_{t\phi}dt d\phi +g_{rr}dr^2 + g_{\phi\phi}d\phi^2,
\end{equation}
since the BTZ black hole is a solution of the Einstein vacuum field equation that meets the above conditions (stationary and axial symmetry), we therefore will use this solution as an input
\begin{eqnarray}
\label{BTZ}
ds^2 &=& -\left(\frac{r^2}{l^2}-M\right)dt^2  - J dtd\phi + \left(\frac{r^2}{l^2}-M+\frac{J^2}{4r^2} \right)^{-1} dr^2 \nonumber \\ && +r^2 d\phi^2 .
\end{eqnarray} 
By obtaining the nonzero matrix $ \mathcal{M} $ in terms of general metric \eqref{ansatz}, combined with the transformation in \eqref{trans} and substituting into of the BTZ black hole metric solution, we obtained the BTZ-Sen solution in string frame
\begin{equation}
\label{gttstring}
\tilde{g_{tt}} = \frac{M+r^2 \Lambda}{\left[\cosh^2\left(\frac{\alpha}{2}\right) +\left(M+r^2 \Lambda\right) \sinh^2\left(\frac{\alpha}{2}\right)\right]^2},
\end{equation}
\begin{equation}
\tilde{g_{rr}} = \frac{4r^2}{a^2 M^2 - 4r^2 (M+r^2 \Lambda)},
\end{equation}
\begin{equation}
\tilde{g_{t\phi}} =-\frac{aM\cosh^2\left(\frac{\alpha}{2}\right)}{2\left[\cosh^2\left(\frac{\alpha}{2}\right) +\left(M+r^2 \Lambda\right) \sinh^2\left(\frac{\alpha}{2}\right)\right]^2},
\end{equation}
\begin{eqnarray}
\label{gpipistring}
\tilde{g_{\phi\phi}}&=& \frac{ \left(\chi_{1} \cosh^2(\alpha)  +\chi_{2} \cosh(\alpha) \right)+\chi_{3}}{\left[\cosh^2\left(\frac{\alpha}{2}\right) +\left(M+r^2 \Lambda\right) \sinh^2\left(\frac{\alpha}{2}\right)\right]^2},
\end{eqnarray}
where
\begin{eqnarray}
&& \chi_{1}=4 r^2 \left(M+\Lambda  r^2+1\right)^2-a^2 M^2 \left(M+\Lambda  r^2+2\right), \nonumber \\ && \chi_{2} =8 r^2+ 2\left(M+\Lambda  r^2\right) \left[a^2 M^2-4 r^2 \left(M+\Lambda 
r^2\right)\right], \nonumber \\ && \chi_{3} =4 r^2 \left(M+\Lambda  r^2-1\right)^2 -a^2 M^2 \left(M+\Lambda  r^2-2\right).
\end{eqnarray}
After obtaining the gravitational field, the corresponding non-gravitational field can be written in the compact form as 
\begin{equation}
\tilde{A_{\mu}dx^{\mu}}= \frac{\sinh(\alpha) \left[(1+M+\Lambda r^2)\tilde{dt}-\frac{aM}{2}\tilde{d\phi}\right]}{\cosh^2\left(\frac{\alpha}{2}\right) +\left(M+r^2 \Lambda\right) \sinh^2\left(\frac{\alpha}{2}\right)},
\end{equation}
\begin{equation}
\tilde{B_{t\phi}}=-\tilde{B_{\phi t}} = \frac{aM \sinh^2\left(\frac{\alpha}{2}\right)}{2\left[\cosh^2\left(\frac{\alpha}{2}\right) +\left(M+r^2 \Lambda\right) \sinh^2\left(\frac{\alpha}{2}\right)\right]},
\end{equation}
and dilaton field
\begin{equation}
\label{dilatonstring}
\tilde{\Phi} = -\ln \bigg|\cosh^2\left(\frac{\alpha}{2}\right) +\left(M+r^2 \Lambda\right) \sinh^2\left(\frac{\alpha}{2}\right)\bigg|.
\end{equation}
With the transformation in \eqref{conformal}, we obtained a black hole solution in Einstein frame which can be written as
\begin{equation}
\label{gtt}
g_{tt}=\frac{(M-b)(M+r^2 \Lambda-b)}{M+b(M-b+r^2\Lambda)}
\end{equation}
\begin{equation}
g_{rr}=\frac{4r^2 \left[M+b(M-b + \Lambda r^2)\right]}{(M-b)\left[a^2 (M-b)^2 + 4r^2 (b-M-r^2\Lambda)\right]},
\end{equation}
\begin{equation}
g_{t\phi}=-\frac{aM (M-b)}{2\left[M+b(M+\Lambda r^2)\right]},
\end{equation}
\begin{eqnarray}
\label{gpipi}
g_{\phi\phi}&=& \frac{a^2 b (M-b)\left[-2M +b(-M+b+r^2\Lambda)\right]}{4 \left[M+b(M-b+r^2 \Lambda)\right]}\nonumber \\ && + \frac{r^2}{(M-b)} \left[M+b(M-b+r^2 \Lambda)\right]
\end{eqnarray}
For non-gravitational field in Einstein frame reads
\begin{eqnarray}
A_{\mu}dx^{\mu} = \frac{Q\left[ \left(1+M-b+r^2\Lambda\right)dt- \frac{a(M-b)}{2}d\phi \right]}{2\left[M+b(M-b+r^2 \Lambda)\right]},
\end{eqnarray}
\begin{equation}
B_{t\phi}=-B_{\phi t} = \frac{a M b (M-b)}{2\left[M+b(M-b+r^2\Lambda)\right]},
\end{equation}
and
\begin{equation}
\label{dilaton}
\Phi=-\frac{1}{2} \ln\bigg|\frac{M+b(M-b+r^2\Lambda)}{M-b}\bigg|.
\end{equation}
Where $ b = \frac{Q^2}{2M} $ and $ a = \frac{J}{M} $ is the rotation parameter. This is a set of solutions from BTZ black holes in heterotic string theory in string and Einstein frame. From the \eqref{gtt}-\eqref{dilaton} solution, if we set $ b = 0 $, then the metric solution will return to the BTZ black hole. For short hand purposes, the we will call this solution as BTZ-Sen Black Hole (BTZ-S BH).

As the author mentioned in the previous section that three-dimensional black holes has similar characteristics to ordinary four-dimensional black holes. A brief explanation of these statement we start by investigating the event horizon in detail and what processes occur in BTZ-S BH. A hypersurface has parametric equation which usually is written as
\begin{equation}
f(x^{0},x^{1},x^{2})=0,
\end{equation}
where $ x^{0}=t $, $ x^{1}=r $, and $ x^{2}=\phi $. The normal vector from the hypersurface can be written as $ n_{\mu}=\frac{\partial f}{\partial x^{\mu}} $. Hypersurface is said to be null if the conditions below are obeyed
\begin{eqnarray}
n^{\mu}n_{\mu}&=&g^{\mu\nu}n_{\mu}n_{\nu}=0, \nonumber \\ && g^{rr}=0.
\end{eqnarray}
We can conclude that $ g^{rr}=0 $ is a way to get the radius of the event horizon. Applying the method to obtain BTZ black holes in heterotic string theory, we get
\begin{eqnarray}
r_{\pm} =l\sqrt{ \frac{(M-b)}{2}\left[1\pm \sqrt{1-\left(\frac{J}{Ml}\right)^2}\right]}.
\end{eqnarray}
Since the black hole that we are looking at is charged, the event horizon of this black hole is parametrized by a mass $ M $, charge $ Q $, cosmological constant $ \Lambda=-\frac{1}{l^2} $, and angular momentum $ J $. To keep the event horizon on this BTZ-S BH still exist, the variables $ l $, $ M $, and $ J $ must obey the following conditions
\begin{eqnarray}
M>b,~ |J|\leq Ml.
\end{eqnarray}
when $ b = \frac{Q^2}{2M} = 0 $, this event horizon will be reduced to the radius in BTZ black holes \cite{Banados:1992gq}. In extreme conditions $ |J| = Ml $, the event horizon radius $ r_{+} $ $ r_{-} $ merges into one with a radius $ r_{e} = l \sqrt {\frac{M-b}{2}} $. Again, this extreme radius is reduced to the extreme radius of the BTZ black hole when $ b $ is turned off. For the case of BTZ-S BH solutions with the condition $ g_{tt}=0 $ 
\begin{equation}
\label{rerg}
r_{sl}=l \sqrt{M-b},
\end{equation}
the value of $ r_{sl} $ is often referred to as the static limit radius; the radius at which the observer will remain stationary even if the black hole is rotating. The ergosphere of BTZ-S BH can be found in the range of event horizon radius and static limit of $ r_{\pm}<r_{erg}<r_{sl} $. To discuss the geometry of this black hole in detail, we start from the event horizon graph

\begin{figure}[htbp]
	\centering\leavevmode
	\epsfysize=5.0cm \epsfbox{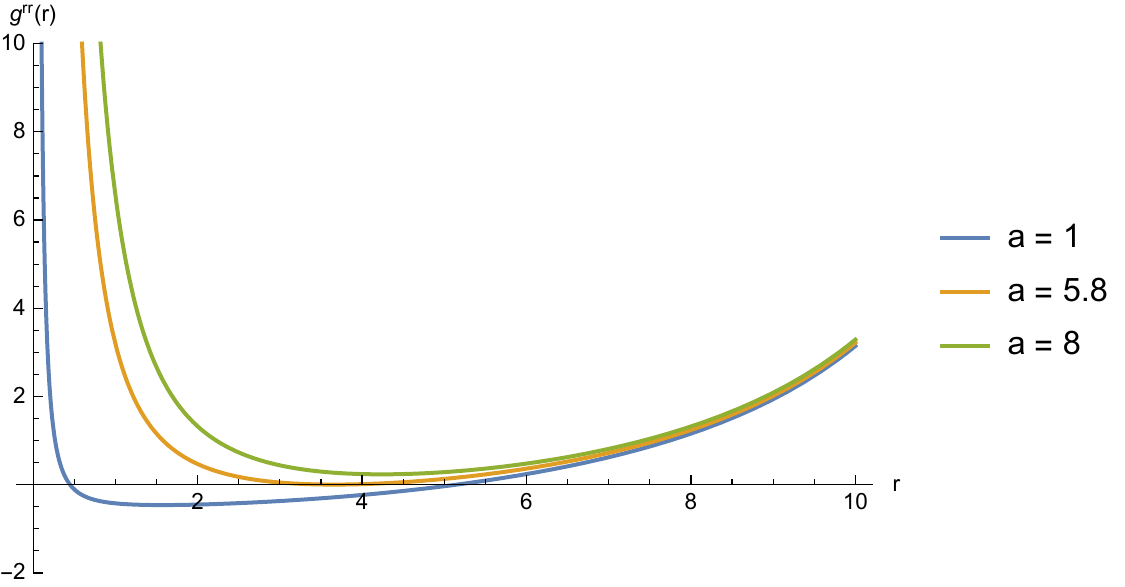}
	\caption {The graph $ g^{rr} $ with respect to the distance $ r $ represents the geometry of the BTZ black hole in heterotic string theory with $ b = -0.2 $, $ M = 1 $, and $ \Lambda=-0.03 $. The metric solution that satisfies the condition $ g^{rr}(r_{+})=0 $ is the radius of the event horizon $r_{+} $. The static limit radius and ergosphere is presented in table~\ref{table:1}.}
	\label{fig:horizona}
\end{figure}

The Fig.~\ref{fig:horizona} is a plot of the metric tensor $ g^{rr} $ with respect to $ r $. We first show the figure with several value of rotation parameters $ a $ to see how much the rotation on BTZ-S BH affects the event horizon(s). From this information, we get that the greater the rotation of the black hole the smaller the distance between the horizon(s). When the rotation parameter shows the value $ a = 5.8 $, the two radii are joined together. After this value, the black hole has a point of singularity that is not covered by event horizons or which can be referred as naked singularities. 
\begin{table}[h!]
	\centering
	\begin{tabular}{||c c c c c ||} 
		\hline
		$ a $ & ~~~~~~~$ r_{-}$ &~~~~~~~~$ r_{+} $  &~~~~~~~~$ r_{sl} $ & \\ [0.5ex] 
		\hline\hline
		1.0 &~ 0.44 & 5.14 & 5.16 &  \\ 
		5.8 &~ 3.65 & 3.65 & 5.16 &  \\
		8.0 &~ - & - & 5.16 & \\ [1ex] 
		\hline
	\end{tabular}
	\caption{Table for exact value of $ r $ from figure~\ref{fig:horizona}. Note that $ r_{ls} $ is a static limit radius. Ergosphere $ \left(r_{erg}\right) $ still exist in the range $ r_{+}\leq r_{erg}\leq r_{ls} $.}
	\label{table:1}
\end{table}

In addition to the event horizon, the three-dimensional black hole has the same characteristics as a four-dimensional black hole, namely ergosphere and static limits. The radius in Eq.~\eqref{rerg} depends only on the mass $ M $, charge $ Q $, and the cosmological constant of $ \Lambda $. We can inferred that, based on the Fig~\ref{fig:horizona} and the table~\ref{table:1}, the rotation parameter $ a $ does not affect the dynamics of the static limit radius. When the naked singularity phenomenon occurs, the static limit radius $ \left(r_{ls}\right) $ will remain.

\begin{figure}[htbp]
	\centering\leavevmode
	\epsfysize=4.7cm \epsfbox{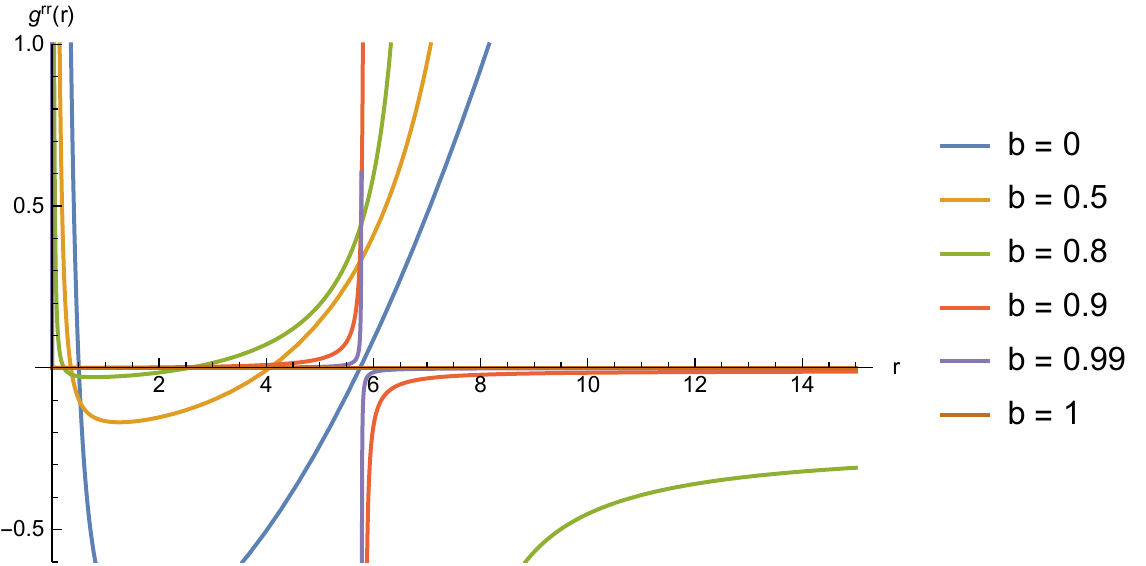}
	\caption {The graph $ g^{rr} $ with respect to $ r $ with $ a = 1 $, $ M = 1 $, and $ \Lambda=-0.03 $. The metric solution that satisfies the condition $ g^{rr}(r_{+})=0 $ is the radius of the event horizon $ r_{+} $. The static limit radius and also the ergosphere is presented in table ~\ref{table:2}.}
	\label{fig:horizonb}
\end{figure}

From the previous story, the variation of the $ a $ parameter does not make the black hole disappear because there are still static surface limits. Now if we make a change on the charge, this black hole will produce a different physical meaning. Therefore, it is very tempting for us to investigate the Fig ~\ref{fig:horizonb}. The figure is a plot that depict the $ g^{rr} $ metric with respect to the distance with the several value of $ b $. It should be reminded again that $ b=\frac{Q^2}{2M} $. In this plot, We want to see the dynamics of event horizons on black holes that begin with $ b = 0 $ to $ b = 1 $. For $ b = 0 $, this black hole has 2 horizon(s) (explaining the geometry of the BTZ black hole). The static limit radius of $ r_{sl} $ from this black hole with a variation of $ 0\leq b\leq 1 $ always coincides with the outer horizon of $ r_{+} $. This is the same as the plot in Fig.~\ref{fig:horizona} when the rotation parameter $ a = 1 $. As the $ b $ parameter increases, it appears that when the outer horizon touches the static limit radius $ \left(M\approx b \right) $, the BTZ-S BH does not exist.

\begin{table}[h!]
	\centering
	\begin{tabular}{||c c c c c ||} 
		\hline
		$ b $ & ~~~~~~~$ r_{-}$ &~~~~~~~~$ r_{+} $  &~~~~~~~~$ r_{sl} $ & \\ [0.5ex] 
		\hline\hline
		0  &~ 0.50 &~ 5.75 &~ 5.77 &~  \\ 
		0.5 &~ 0.35 &~ 4.06 &~ 4.08 &~  \\
		0.8 &~ 0.22 &~ 2.57 &~ 2.58 &~  \\
		0.9 &~ 0.15 &~ 1.81 &~ 1.82 &~  \\
		0.99&~ 0.05 &~ 0.57 &~ 0.57 &~  \\
		1.0 &~ 0    &~ 0    & 0  &~  \\ [1ex] 
		\hline
	\end{tabular}
	\caption{Exact value of $ r $ from Fig.~\ref{fig:horizonb}. Note that $ r_{ls} $ is a static limit radius. The ergosphere $ \left(r_{erg}\right) $ exist in the range of $ r_{+}\leq r_{erg}\leq r_{ls} $.}
	\label{table:2}
\end{table}

Now look at a particle in the ergosphere. Particles that we can still analyze must move in the timelike and null geodesics $(ds^2 \leq 0) $. With $ r=\phi = $ constant, we can get
\begin{equation}
ds^2 = \frac{(M-b)(M+r^2 \Lambda-b)}{M+b(M-b+r^2\Lambda)} dt^2,
\end{equation}
Since we want to see particles moving in the ergosphere $ (r <r_{1}) $, we get the conditions $ \frac{r^2}{l^2}<(M-b) $. The consequences of this relationship are $ds^2> 0 $; Particles that stand still in the ergosphere tend to move in a spacelike manner. So other terms in the tensor metric must contribute and the particle must change position. Fortunately, the case above corresponds to a black hole BTZ $ ds^2 =\left(M-\frac{r^2}{l^2}\right) dt^2 $, with the condition $ \frac{r^2}{l^2}<M $.

The limitation above requires us to change the condition of the particles. If the particle moves in the direction of $ \phi $ with $ t $ and $ r $ constant, then the particle has an angular velocity around BTZ-S BH that satisfies the equation
\begin{equation}
\Omega^2 + 2 \frac{g_{t\phi}}{g_{\phi\phi}}~ \Omega + \frac{g_{tt}}{g_{\phi\phi}}=0,
\end{equation}
Thus, we find the angular velocity
\begin{equation}
\Omega_{max,~min} = -\frac{g_{t\phi}}{g_{\phi\phi}} \pm \sqrt{\left(\frac{g_{t\phi}}{g_{\phi\phi}}\right)^2 - \frac{g_{tt}}{g_{\phi\phi}}}.
\end{equation}
The angular velocity obtained above depends only on the distance. In general, because $ g_{tt}<0~(\Omega_{min}<0) $ so the observer can see the particle rotating in the opposite direction to the rotation of the black hole. When the particle is near the static surface limit of $ g_{tt} \rightarrow 0 $, the angular velocity of the particle has two values, namely $ \Omega_{min}=0 $ and $ \Omega_{max}=-2 \frac{g_{t\phi}}{g_{\phi\phi}} $. That is, the particle (at $ \Omega_{min} $), is remain at rest instead of rotating in the opposite direction of the black hole rotation. Thus, the maximum angular velocity of $ \Omega_{max} $ can be explicitly stated
\begin{equation}
\label{kecepatansudut}
\Omega_{max}= \frac{2M \left[M+b(M-b+\Lambda r^2)\right]}{ab\left[M+b(M+\Lambda r^2)\right]\left[b(b-M+\Lambda r^2)-2M\right](1+\chi_{4}^2) },
\end{equation}
where $ \chi_{4}= \frac{2r\left[b(M-b+\Lambda r^2)-2M\right]}{a(M-b)\sqrt{b^2(b-M+\Lambda r^2)-2Mb}}$. To ensure the analysis is still on the right track, we can prove that if the charge term $ b $ is turned off, then the maximum angular velocity above belongs to the black hole BTZ $ \left(\Omega_{max} = \frac{a}{2 M r^2} \right) $. In addition, Eq.~\eqref{kecepatansudut} also satisfies physical conditions. At a very large distance $(r\rightarrow \infty) $, the angular velocity is zero. This indicates that the particles that are very far from the black hole do not feel the frame dragging effects at all. This observation again adds to our belief that the family solutions of three-dimensional black hole has the same characteristics as the family solutions of four-dimensional black hole. In the case of Kerr black hole, zero angular velocity at infinity is caused by an asymptotically flat space-time background. However, the angular velocity with asymptotically-flat anti-de Sitter also has the same result. From the previous statement, we can conclude that the particle will not feel the frame dragging effect no matter what the asymptotic background is. The angular velocity that gets smaller when the distance is farther away from the object also obeys the conservation law of angular momentum. Moreover, particles that enter the ergosphere area will rotate in the direction of rotation of the BTZSBH around
\begin{equation}
L = 2 \pi r_{+}= 2\pi l\sqrt{ \frac{(M-b)}{2}\left[1+ \sqrt{1-\left(\frac{J}{Ml}\right)^2}\right]}.
\end{equation}
The formula above is the same as the perimeter of a circle since a 3-dimensional (2+1) black hole has a 2-dimensional surface. Finally, when the charge contribution is switched off, the perimeter of the BTZ black hole is again obtained.

With the end of this section, we already know the geometry and characteristics of BTZSBH. Starting from solutions on the string and Einstein framework, event horizon, ergosphere, static radius limit, angular velocity, and circumference of BTZ-S BH. In conclusion, BTZ black holes in the low energy heterotic string theory still exist in space-time.

\section{Conclusion}
In this paper, we construct the classical exact solution of a three-dimensional black hole in low-energy heterotic string theory (BTZ-Sen balack hole). The exact solution we obtain are BTZ black hole \cite{Banados:1992gq,Banados:1992wn} with Maxwell $ F_{\mu\nu} $, nontrivial dilaton $ \phi $, and 3-form field $ H_{\mu\nu\rho} $. The Hassan-Sen transformation is the path we choose to obtain the metric solution. This method requires us to have seed solutions that obey the stationary and axisymmetric conditions. Since this novelty involves two important concepts, we structured this paper into two sections; (i) BTZ black hole as a review, and (ii) BTZ Black Hole in the low energy heterotic string theory (BTZ-Sen). In our discussion on BTZ black hole in the low energy heterotic string theory, the solution we obtain as a result of the Hassan-Sen transformation starts from the string frame. The metric solution is still paramaterized by the mass $ M $ and angular momentum $ J $. We employ Sen \cite{Sen:1992ua} combined with the method in~\cite{Magnon:1985sc,Brown:1992br,Kastor:2008xb} to obtain the conserved mass, angular momentum, and charge since the object obey the lower derivative gravity $ (\Lambda \neq 0) $. Expressing the conserved quantity in terms of ADM variable, we obtain the charge case after conformal transformation to Einstein frame~\eqref{gtt}-\eqref{gpipi}. When $ b $ is switched off, the metric solution automatically reduces to the well known BTZ black hole.

From our investigation of the event horizons, to keep the black hole exist the cosmological constant parameters $ l $, mass $ M $, angular momentum $ J $, and charge $ b $ must oeby the conditions $ M>b $ and $ |J| \leq Ml $. Under these conditions, we obtain that the number of horizons will decrease as the  rotation parameter $ a $ increases. A different story arises when the parameter $ b = \frac {Q^2}{2M} $ being varied. For $ b = 0 $, the event horizon characteristics are the same as BTZ black holes. When the $ b $ parameter increases, the event horizon shrinks until the $ b $ parameter equals to the value of $ M $, a black hole does not exist. For another aspect, particles that stand still in the ergosphere have  space-like conditions $(ds^2> 0) $ so they must orbit in the same direction of rotation as the black hole. The obtained angular velocity is reduced to BTZ angular velocity when the $ b $ variable is turned off. In the far region, the particle does not feel the frame dragging effect of the rotation of the black hole no matter what the asymptotic background is.

With this background solution, it is tempting to investigate the phenomenological aspect from the solution. One can obtain the thermodynamical properties and the first-law of BTZ-S BH \cite{novelty1}. Other investigations are also of interest such as geodesics, deflection, and photon sphere around BTZ-S BH using the Hamilton-Jacobi equation \cite{novelty2}. Since the black holes also obeys quantum phenomenology, we can also look for massless scalar wave scattering due to the appearance of BTZ-S BH. This case can be obtained by using the massless Klein-Gordon equation to obtain the field $ \Phi (t,r,\phi) $. By using JWKB approximation to get the phase shift of the effective potential that exists from the interaction, and the eikonal approximation to obtain the scattering amplitude, the scattering cross-section can be obtained from the absolute square of $ f(\theta) $ \cite{novelty3}. We leave these three novelties for our further investigation. 

\section{Acknowledgement}

We thank Ilham Prasetyo for the insight on Hamiltonian and Lagrangian in curved space-time. We also thank Haryanto M. Siahaan, Handhika S. Ramadhan, and A Sayyidina for enlightening discussions. This work is partially funded by the Q1Q2 grants from Universitas Indonesia under the contract No.~NKB-0270/UN2.R3.1/HKP.05.00/2019.

\end{document}